# Implications of Managing Health related Records and Relevant Information Systems within Intergovernmental Agencies

**Adam Davies, Roberto Bergami\* and Shah Jahan Miah**

College of Business, Victoria University Melbourne, Australia
(\* & Visiting Professor, University of South Bohemia in Ceske Budejovice, Czech Republic)
Email:  adam.davies1@live.vu.edu.au;
roberto.bergami@vu.edu.au;
shah.miah@vu.edu.au

## Abstract
The implications of intergovernmental agencies may forever change the way in which governments provide common services within a federated Australia. As governments seek to reduce duplication and inconsistencies across state and territory borders, intergovernmental agencies are faced with the challenge of managing health related records under differing laws and with lack of clarity on ownership of each record. As records and cost of records increases within these entities we examine if an intergovernmental agencies can  ever  dispose  of  a record or does the legal frameworks for these agencies mean that the information systems need to evolve to support these new and emerging entities? This paper will examine the introduction of intergovernmental agencies and the challenges of managing health related records and relevant information systems within these agencies, to explorer how recent legal precedents or the concept of information citizenship may provide guidance on how to manage transient records and cloud services, while also mitigating the impacts of data sovereignty.

## Introduction

The introduction of intergovernmental agencies (IGAs), by the Council of Australian Government (COAG), aimed at reducing "duplication and inconsistencies across state and territory borders" (NHVR 2013) has profound implications on the future of  records management within Australia. IGAs being "established by state and territory governments through the introduction of consistent legislation in all jurisdictions" (Health 2015) do not result in an entity, reflective of that of a Commonwealth agency, thereby reducing the flexibility in which they can manage records, but result in an entity for which the collective legislative framework can be structured to allow the entity to  predetermine  which combination of records legislation will apply to themselves.

Recent developments, such as the bilateral agreement between Victoria  and  New  South Wales on the legal profession uniform framework highlight the potential of IGAs to select which records legislation they desire to comply. Evidence of such a selection can be found in the Victorian Legal Profession Uniform Law Application Act (2014), which in section 5 notes that "the following Acts of this jurisdiction do not apply to the Legal Profession Uniform Law (2014) or to instruments made under that law, (a) the Audit Act 1994  …  (f)  the  Public Records Act 1973" (p.4). This is a precedent to reduce the legislative burden placed on the administration of legal service with enormous implications for other IGAs, such as the Australian Children's Education & Care Quality Authority (ACECQA), National Heavy

Vehicle Regulator (NHVR) and the Australian Health Practitioner Regulation Agency (AHPRA) to review the structure in which they were created.

Research on the implications for records management arising from IGAs is noticeably limited. Literature on IGAs tends to focus on such topics as the regulations they administer (King 2013): the history of their creation (McIntosh 2011); or the progress they have made (Marty 2012). Researchers need to draw parallels between records management challenges facing Non-Government Organisations (NGOs) operating over similar jurisdictional boundaries, in seeking to support further studies, with parallels found in such topics as data sovereignty (Peterson et.al. 2011); cloud computing (Griebel et.al 2015); and audit and information security (Bendovschi & Ionescu 2015).

Health related IGAs currently form the greatest concentration across Australia The implications to their management of health related records and information systems is further challenged by the, now eight, state and territory specific national bodies general disposal authorities developed by the Council of Australasian Archives and Records Authorities (CAARA) (Council of Australasian Archives and Records Authorities 2008). These authorities cover the retention and disposal of the administrative subset of records held by an IGA. The challenge identified by this investigation is that each State and Territory had to endorse its own version of the authority under their legislative framework, raising such questions as: i) how do IGAs determine ownership of a record?; ii) how does the legislation accommodate the transfer of records across jurisdictions? and iii) can an IGA ever dispose of a record, when questions may persist around the appropriate use of any of the disposal authorities?.

The paper is structured to provide the reader with an overview of the 'as is' state of records management within IGAs, and also to support the selection of the health industry as the focus of this research. Firstly, this paper presents a background history surrounding IGAs, and the progress to date in developing appropriate retention and disposal authorities (RDA) to support the common obligations of selecting disposal, as this is used in the analysis. The methodology section walks through the selection of disposal as the records management component to focus on and also to provide an overview of current IGAs operating in Australia, justifying the choice of the health industry as the exemplar for this research. The discussion section reviews the implications of managing of records and their information systems across jurisdictional boundary and considers the potential of granting information citizenship as method of formalising data sovereignty. The concluding section summaries the implications of managing health related records and their information systems within an IGA, identifying opportunities and challenges of such legislative frameworks, while exploring the concept of information citizenship as bridging framework, from which broader theoretical and practical contributions can be made to assist these new and emerging entities.

## Background

Research into the management of records within an IGA is in its infancy, as these new and evolving organisational structures emerged recently through changes to legislative frameworks. In seeking to build compliant information systems, managers within these organisations face the challenges observed by Yusof, et al (1998) in that "records management, both as a profession and as a discipline, is relatively new" (p. 13). This observation provides insight into to challenges facing the information management professions, as they seek to find their place within, a largely technology driven, information landscape, where organisations are grappling with such topics as data sovereignty (Irion, 2012); cloud computing (Vaile, 2014); and centralisation (De Filippi & McCarthy, 2012). As technology across all fields drives changes in the quantity and quality of records and related systems, researchers, such as Cummings & Findlay (2010), are already asking: "Are we tipping into a digital oblivion, a period of extensive data loss, a period where records managers will collectively fail in their quest to manage digital records?" (p. 267).

In exploring the implications of technology on health organisations, Deokar & Sarnikar (2014) observed that "in the past decade, healthcare organizations have greatly accelerated their investments in information technology" (p.1). This investment in technology and subsequent increase in creation of electronic records, is challenging the supremacy of paper, one that Lappin (2010) suggested was so strong that all "existing theory and practice was predicated on the assumption that organizations were keeping records in paper form" (pp. 253-254). In their potential to streamline the collection and dissemination of information, information systems are enabling new concepts, as identified by Gunter & Terry (2005) in stating "the electronic health record (EHR) is an evolving concept … [where] … considerable uncertainty exists regarding the costs associated with electronically mediated health initiatives" (Gunter & Terry, 2005, p.1 &7).

The concept of the IGA was defined by COAG (2009) in the National Partnership Agreement on the Quality Agenda for Early Childhood Education and Care as a "new body which will be established to guide the implementation and management of the new integrated [framework]" (p.6), providing the closest governmental equivalent to that of an Australia wide NGO. The similarity between government and NGOs arise from the requirement, that each needs to comply with legislative frameworks, of each jurisdiction in which they operate. This requirement is a new concept in government, as previously government entities were either State, Territory or Commonwealth entities only needing to comply with one appropriate set of legislation. Intergovernmental agreements, such as the national registration and accreditation scheme for the health professions (NRAS 2008), however, require that "States and Territories undertake to use their best endeavours to submit to their respective Parliaments whatever Bill or Bills that have the effect of achieving a national scheme" (NRAS 2008, p. 4), thus creating an entity for which compliance is required under one or more appropriate sets of legislation.

The national nature of IGAs does not require entities to operate in each and every state and territory of Australia, as does the Australian Health Practitioner Regulation Agency (AHPRA). In fact, many IGAs such as NHVR, ACECQA, NHPA and the Independent Hospital Pricing Authority (IHPA), are either not represented by all State and Territories, or a collective of Commonwealth and non-Commonwealth entities working together to administer a national framework. The commonality between all of these IGAs is that, in their legislative frameworks, no sovereign entity has excluded their obligations to manage records, under their records legislation. CAARA (2015), the peak body for records management within Australia "comprising the head of the government archives authorities of the Commonwealth of Australia, New Zealand and each of the Australian States and Territories", defines in its Policy 11 – Guidelines for the Treatment of Records of Inter-Governmental Agencies, that an IGA is "a joint administrative agency established by more than one government whether at Commonwealth, State or Territory level to conduct business of common interest."

As IGAs, such as those mentioned above, seek to comply with all of their legislated record obligations, information systems, assisting in the capture, control and storage of records, may be further challenged, as information and record stores are located in different jurisdictions to that of which the creator resides. In its use to provide additional context to each record, metadata may require the inclusion of citizen-like properties, allowing systems and operators to record the source jurisdiction, which may then be used to inform the overall ownership of the record and, therefore, which legislation is appropriate for its management. As IGAs seek to manage their costs associated with record storage, they are restricted in their management options to either disposal of records under an RDA, or transfer of records to an archival institution. Whilst these options appear reasonably straight forward for a non-IGA, they are further complicated for multi-jurisdictional IGAs, as CAARAs Policy 11 requires that "no archival institution is to claim an IGA's records until consultation between interested archival institutions has occurred" (p.1). In the event consultation has not concluded, IGAs are left with the prospect of managing transient information stores and ever increasing storage costs.

Authorisation to disposal of information is granted via an RDA. Disposal authorities come in two main forms: general retention and disposal authorities (GDA), covering records of a common function, and specific RDAs, focused on specific records and records classes of a particular organisation's functions. Although these authorities are known by different names in some States and Territories, their function and application remains the same. The concept of a GDA is not new, with "General Records Disposal Schedules for administrative, personnel and financial records [existing] since 1978" (Robinson 1997, p. 298). Retention and disposal authorities are granted their legal status through the appropriate records Acts of their specific jurisdiction, as shown in Table 1.

| Jurisdiction | Title |
|---|---|
| Australian Capital Territory | Territory Records Act 2002 |
| Commonwealth of Australia | Archives Act 1983 |
| New South Wales | State Records Act 1998 |
| Northern Territory | Information Act 2003 |
| Queensland | Public Records Act 2002 |
| South Australia | State Records Act 1997 |
| Tasmania | Archives Act 1983 |
| Victoria | Public Records Act 1973 |
| Western Australia | State Records Act 2000 |

*Table 1 - List of archival legislation across Australia*

The "General Retention and Disposal Authority (GDA) for Administrative Records of National Bodies [was] approved by the Council of Australasian Archives and Records Authorities (CAARA) on 18 October 2013" (State Record Office 2014, p. 6). Referring to IGAs as national bodies, this authority allows for the disposal of records considered to be administrative in nature, while also restricting IGAs from disposing of records "created by national bodies' predecessor agencies … [or] … functions that are unique to a national body(s)" (State Record Office 2014, p. 6). The challenge facing IGAs in using GDAs is that whilst CAARA may be the peak body for records management across Australia, it lacks legislative power to enforce such GDAs. The outcome was that each jurisdiction had to enact its own GDA version, resulting in eight disposal authorities aiming to be identical. A summary of the State/Territory GDA authorities as at 30 June 2015 is shown at Table 2.

| Jurisdiction | Title | Authority | Issued |
|---|---|---|---|
| Australian Capital Territory | Records Disposal Schedule - National Bodies Administrative Records | NI2015—34 | 2015 |
| New South Wales | GA43 General authority for national bodies | GA43 | 2014 |
| Northern Territory | General Records Disposal Schedule Administrative Records of National Bodies | No. 2013/9 | 2013 |
| Queensland | National Bodies General Retention and Disposal Schedule for Administrative Records | QDAN 711 v.2 | 2014 |
| South Australia | Administrative Records of National Bodies | GDS No.34 | 2014 |
| Tasmania | Disposal Schedule for functional administrative | 2015: | 2015 |

|  | records of Inter-Governmental Agencies | DA2437 |  |
|---|---|---|---|
| Victoria | Retention & Disposal Authority for Administrative Records of National Bodies | PROS 13/07 | 2013 |
| Western Australia | General retention and disposal authority for administrative records of nation bodies. | 2014004 | 2014 |

*Table 2 - Endorsed equivalent of the national bodies' authority in each State and Territory*

As IGAs seek to manage records, information systems and storage costs, has the federated legal framework within Australia rendered the notion of national services unworkable, or do information systems and their supporting frameworks need to evolve in support of these organisational structures? This paper aims to provide answers to this question following the methodology section.

## Methodology

Due to size limitations, the paper focuses only on the practical challenges of managing record disposal within the health industry. As records management provides a common set of requirements to baseline across all organisations, government or NGO, it serves as a mechanism for evaluating implications of any change in management of records and their systems, by investigating legal frameworks of different types of organisations and noting their failures and/or successes.

Potential differences in the implications of managing records and their systems within IGAs resulting from differing legislative frameworks are acknowledged. Therefore, this analysis will take a meta-view of legislative frameworks governing specific IGAs, as formed by COAG intergovernmental agreements. The selection of disposal, as the records management principle to be investigated, was based on evaluation of the South Australia's State Record Office, assessment of key components of the records management lifecycle, as shown in Table 3.

| Component | Summary |
|---|---|
| Creation | Official records are created as a direct consequence of the conduct of the business of government. The records:<br>• provide proof that certain actions or events occurred<br>• meet specific legislative requirements concerning the creation of records<br>• enable the agency to see what has happened in the past and act as an information source to guide future actions (State Record Office, 2012b, p.2) |
| Capture | The intent of capturing records into a recordkeeping system is to:<br>• establish a relationship between the record, creator and business context<br>• place the record and its relationship within a recordkeeping system<br>• link the record to other records (State Record Office, 2012b, p.2) |
| Control | Control of official records is maintained through classification and application of other metadata. (State Record Office, 2012b, p2) |
| Storage | Implement recordkeeping/business systems and storage facilities that are protected from unauthorised access, intentional illegal destruction or theft, and from damage (State Record Office, 2012c, p.1) |
| Access | The security of records is essential to ensuring their reliability, integrity and evidential value. It is important that agencies understand the sensitivity of the records they hold, as this is key to correctly identifying the security classifications and measures which should be applied to systems, physical locations and staff members. (State Record Office, 2012c, p.1) |

| Component | Summary |
|---|---|
| Disposal | The benefits of a well-managed disposal program are many. Knowing how long they are required to keep particular records allows agencies to manage storage costs. Being able to explain why records are no longer held enables agencies to demonstrate compliance with legislation such as the Freedom of Information Act 1991. Critically, disposal programs allow agencies to identify those records which are of long term value and form part of the corporate memory of the government and the collective memory of society (State Record Office, 2012d, p.1) |

*Table 3 - Components of the records management lifecycle*

It was determined that disposal was most appropriate for this study as in order for any IGA to dispose of records, it must have previously already undertaken the process of creation, capture, control, storage and access. Each IGA, through its own unique legislative framework, can determine how these prior processes and which systems are used to support them are undertaken. However, as each IGA then is required to return to a common disposal authority the process of disposal is the common baseline. The health industry was selected to complement the principle of disposal, following an initial evaluation of IGAs resulting from intergovernmental agreement, as at the 30th of June 2015, as shown in Table 4.

| Intergovernmental agreement | Field | Resulting entity | Formed |
|---|---|---|---|
| Intergovernmental agreement for a national registration and accreditation scheme for the health professions | Health | Australia Health Practitioner Regulation Agency | 2010 |
| Intergovernmental agreement for an electronic conveyancing national law | Finance | National Electronic Conveyancing Office | 2005 |
| Intergovernmental agreement on heavy vehicle regulatory reform | Transport | National Heavy Vehicle Regulator | 2013 |
| National partnership agreement on the national quality agenda for early childhood education and care | Education | Australian Children's Education & Care Quality Authority | 2012 |
| National health reform agreement | Health | Independent Hospital Pricing Authority | 2011 |
| National health reform agreement | Health | National Health Funding Body | 2012 |
| National health reform agreement | Health | National Health Performance Authority | 2014 |
| National health reform agreement | Health | Australian Commission on Safety and Quality in Health Care | 2006 |
| Intergovernmental agreement for regulatory and operational reform in occupational health and safety | Workplace | Safe Work Australia | 2009 |

*Table 4 - Initial list of national bodies operating within Australia as at 30 June 2015*

It can be observed from Table 4, the majority of entities operate within the health sector, and this supports and validates the choice of this industry as an exemplar for this investigation. It may be the findings are applicable to other sectors and this is considered later in the paper. The next section discusses the implications of managing health related records and their systems within an IGA, exploring opportunities and challenges facing information managers, as they navigate their specific legislative framework and also further exploring the concept of information citizenship as framework to managing transient information, similar to how governments manage transient people.

## Discussion

Records managements is facing a new frontier as IGAs, tasked with the management of health information and knowledge, challenge the way in which records are managed, controls used, and ownership determined. As health organisations implement e-health solutions, which are tailored to support privacy, security and management of electronic health records, greater numbers of records management professionals experienced in managing large scale electronic collections across multiple jurisdictions are required. To gain the experience required to support these entities, records and information managers must firstly determine who owns the records, as this will inform which legal framework needs to be complied with. The consequences of not determining ownership include "failure to clarify issues surrounding the legal ownership of records, and the information they contain, in outsourcing agreements and contracts can severely restrict the business capabilities of the contractor and expose the organization to considerable risks" (Commission of Western Australia 2002, p.6).

Ownership of a paper based record is determined by the type and location of the government agency that creates or receives such record. This is reasonably straight forward as agencies in New South Wales receiving correspondence for the New South Wales Government may easily claim that record is governed by New South Wales law. Determining ownership of electronic records, however, is not as straightforward. Electronic records received by an entity in New South Wales may actually never enter that State as information systems or cloud services may physically store those records on a server in another jurisdiction. The internet, cloud and many outsourced services rely on this notion of storing information outside of the physical boundaries of organisations that procure them. This architecture requires organisations, especially IGAs, to assess the impact of data sovereignty on the ownership of records as the physical location of the server may be the determining factor in identifying which legislative framework needs to be applied.

The implication of managing health related records within an IGA, is further complicated due to the transient nature of society, with patients and practitioners moving between jurisdictions, creating records, amending others and including new content potentially owned by the jurisdiction in which it was created. Health related records may in fact contain a range of content collectively owned in part by each and every jurisdiction across Australia. In architecting or procuring information systems IGAs may in the future require their systems to capture and retain key metadata identifying the jurisdiction in which the record was first created. The concept of capturing jurisdictional metadata, as form of identifying where records where created, may be similar to how governments capture citizenship information to determine the nationality of human beings, which can be used to subsequently determine which legal framework should be applied in cases of legal uncertainty. The concept of information citizenship, as a framework for managing records in a global environment cannot be explored in sufficient depth within this paper due to size limitations, but it is a concept worthy of further separate inquiry.

In investigating some of the legal differences health IGAs face in managing health related records across some of Australia's most populated jurisdictions, it is evident the requirements of the Victorian Public Record Act 1973 being "a health-service provider must not delete or dispose of health information unless it is permitted to do so under a current [Public Record Office Victoria] PROV Records Authority that falls under the PR Act." (Public Record Office Victoria 2003, p.2) are not compatible with those of the New South Wales Public Record Office (2015) which states "destruction of State records as part of a program of authorised records disposal in accordance with Part 3 of the State Records Act 1998". The incompatibility of requirements between the Victorian and New South Wales Public Record Acts further strengthens the need for health IGAs and their information systems to be able to identify the owner of each record, so they may implement appropriate disposal programs.

In striving to maintain currency in a technology driven world, health related IGAs are looking for new and innovative way of delivering health services, supported by "electronic records [that] are shaping up to be the future of health care" (O'Sullivan et.al, 2011, p.179). As health information becomes more digital, ownership is not the only challenge facing IGAs, with health related "records contain[ing] highly sensitive health and legal information, so ensuring confidentiality is a paramount concern" (Bismark et.al, 2015, p.2). The combination of managing confidentiality alongside uncertainty of legal precedent, may be a contributing factor to lack of literature or market expertise in managing these new governmental entities. Managers, however, should not be disheartened by challenges facing them, as concepts such as information citizenship or the precedent set by the Victorian Legal Profession Uniform Law Application Act (2014) provide potential contributions to broaden theory and practices of managing IGAs.

The Victorian Legal Profession Uniform Law Application Act (2014) will have resounding effects of the future formation of any IGA, as it is the first IGA to select to exclude the resulting entity from needing to comply with the Public Record Act of the jurisdiction in which it operates. This precedent raises questions around who would own any record created by the resulting entity. If the record was created on a server in Victoria, data sovereignty would require the record to be Victorian, however, the legal framework excludes the entity from needing to comply with the appropriate laws in Victoria. The record may in fact be considered state-less, being similar to a human without any nationality. The issue facing records managers is that while a record may be state-less it is impossible to manage these records, as management of record requires a framework to manage from, and the framework is determined by the owner or nationality of the record.

Without the identification of a record owner, IGAs will never have the confidence to dispose of any health related records, as under such circumstances no disposal program would be legally defensible. Without a disposal program IGAs will be faced with the prospect of perpetual record retention, resulting in either of two possible outcomes: i). IGAs will need to continually increase the cost of services provided, or ii) if these costs cannot be passed on to clients, the IGAs will eventually become insolvent. The conclusion focuses on the potential of both the concept of information citizenship as well as the precedent of the Victorian Legal Profession Uniform Law Application Act (2014), with a view to identify information systems and architectures implications as new health related IGAs challenge the way health information services are traditionally provided.

## Conclusion

The introduction of IGAs has forever changed the perception on how records are to be managed within multi-jurisdictional organisations. The structure of an IGA, being reflective of large academic institutions or corporate organisations, means that relevance of any research findings into the management of transient information flows, can be applied not only to IGAs but also to many organisations within Australia and across the globe. As technology enhances the opportunities for records to be created, researchers and information system professionals alike may need to reconsider how records are created and also which metadata is required as a mandatory baseline to support the ongoing management of each and every record.

The precedent contained in the Victorian Legal Profession Uniform Law Application Act (2014) may be one option for IGAs to mitigate the challenges of creating and capturing records across jurisdictions. However, if this precedent is not applicable to all IGAs or any reflective organisation then information systems and systems architectures may need to evolve, to support these new and expansive organisation types. The potential of health related IGAs to undertake disposal is but one of many potential records management investigations that could be undertaken to assist multi-jurisdictional organisations manage their risk and reduce their costs.

Healthcare and use of electronic health records are evolving to meet demands placed on them by society. Organisations that develop information systems solutions will likewise need to evolve to successfully meet the demands of these new health focused entities. The ability to identify the owner of a record is, therefore, paramount for without a clearly defined owner, records managers are unable to determine which legal framework needs to be applied. The concept of capturing additional metadata as a mandatory baseline for all information may be one avenue information systems developers can pursue to allow organisations to set the citizenship for information created under their control.

In a world where information is identified as owning or belonging to a jurisdiction, the challenges of managing records becomes a little easier, for with the owner of each record being identified, it is then possible to develop information systems to apply records management practices across all records with the same owner, regardless of where they are stored. The answers to the challenges facing health related IGAs may well be in the technology used to capture and manage records. The future of electronic health records may result in records obtaining passport-like details, at least until the federated legal structure of Australia becomes as little less complicated.

# REFERENCES

Bendovschi, A. C., & Ionescu, B. D. (2015). The Gap between Cloud Computing Technology and the Audit and Information Security. Audit Financiar, vol.13, p.125.

Bismark, M. M., Fletcher, M., Spittal, M. J., & Studdert, D. M. (2015). A step towards evidence-based regulation of health practitioners. Australian Health Review.pp.1-3

Bowen, G. A. (2009). Document analysis as a qualitative research method. Qualitative research journal, vol.9, pp.27-40.

Council of Australasian Archives and Records Authorities (2008), CAARA Policy 11 – Guidelines for the Treatment of Records of Inter-Governmental Agencies, viewed at http://www.caara.org.au/index.php/policy-statements/guidelines-for-the-treatment-of-records-of-inter-governmental-agencies/ on 21/07/2015.

Cumming, K & Findlay, C (2010), 'Digital recordkeeping: are we at a tipping point?', Records Management Journal, vol.20, no.3, pp.265-78.

De Filippi, P., & McCarthy, S. (2012). Cloud Computing: Centralization and Data Sovereignty. European Journal of Law and Technology, vol.3, no.2, pp.1-25

Deokar, A. V., & Sarnikar, S. (2014). Understanding process change management in electronic health record implementations. Information Systems and e-Business Management, pp.1-34.

Griebel, L., Prokosch, H. U., Köpcke, F., Toddenroth, D., Christoph, J., Leb, I., & Sedlmayr, M. (2015). A scoping review of cloud computing in healthcare. BMC medical informatics and decision making, vol.15, no.1, p.17.

Gunter, T. D., & Terry, N. P. (2005). The emergence of national electronic health record architectures in the United States and Australia: models, costs, and questions. Journal of Medical Internet Research, vol.7, no.1, pp.1-17.

Health (2015), Australian Government, Department of Health, viewed at http://www.health.gov.au/internet/main/publishing.nsf/Content/work-nras

Irion, K. (2012). Government cloud computing and national data sovereignty. Policy & Internet, vol.4, no.3-4, pp.40-71.

King, E. (2013). Australia's youngest of citizens. Developing Practice: The Child, Youth and Family Work Journal, vol.36, p.52.

Lappin, J. (2010). What will be the next records management orthodoxy? Records Management Journal, vol.20, no.3, pp.252-264.

Legal Profession Uniform Law Application Act (Vic), (2014) State Government of Victoria, viewed at http://www.legislation.vic.gov.au/domino/Web_Notes/LDMS/LTObject_Store/ltobjst9.nsf/DDE300B846EED9C7CA257616000A3571/E737D364AFDF26E5CA257E2F00139C2B/$FILE/14-17aa003%20authorised.pdf

Marty, S. (2012) From the PBA: AHPRA turns two [online].AJP: The Australian Journal of Pharmacy, Vol. 93, No. 1107, p.1.

McIntosh, R. (2014), NHVR - A United Front Essential [online].Diesel, Vol. 11, No. 2, p.1

NHVR (2013), National Heavy Vehicle Regulator, viewed at https://www.nhvr.gov.au/files/fact-sheet-2-national-heavy-vehicle-accreditation-scheme.pdf

O'Sullivan, T. A., Billing, N. A. & Stokes, D. (2011). Just what the doctor ordered: Moving forward with electronic health records. Nutrition & Dietetics, vol.68, pp.179-184.

Peterson, Z. N., Gondree, M., & Beverly, R. (2011,). A position paper on data sovereignty: the importance of geolocation data in the cloud. In Proceedings of the 8th USENIX conference on Networked systems design and implementation.

Public Record Office Victoria, (2003), PROA5, Advice 5 Health Records and Public Records Advice, viewed at http://prov.vic.gov.au/wp-content/uploads/2011/05/PROVRMAdvice5.pdf

Robinson, C. (1997). Records control and disposal using functional analysis. Archives and Manuscripts, vol.25, pp.288-303.

State Records Act (1998), State Government of New South Wales, State Records Act 1998 No 17, viewed at http://www5.austlii.edu.au/au/legis/nsw/num_act/sra1998n17183.pdf

State Records Commission (2002), Western Australian Government, Standard 6: Outsourcing, viewed at http://www.sro.wa.gov.au/sites/default/files/src-standard6.pdf

State Record Office (2012a), South Australian Government, General Disposal Authorities http://government.archives.sa.gov.au/content/general-disposal-schedules#GDS24

State Record Office (2012b), South Australian Government, RK050 - ARM - Creation, Capture and Control, viewed athttp://government.archives.sa.gov.au/sites/default/files/20131106%20Adequate%20Records%20Management%20-%20Creation%2C%20Capture%20%26%20Control%20Final%20V1_Copy.pdf

State Record Office (2012c), South Australian Government, RK052 - ARM – Security and Accessibility V1, viewed at http://government.archives.sa.gov.au/sites/default/files/20131106%20Adequate%20Records%20Management%20-%20Security%20%26%20Accessibility%20Final%20V1_Copy.pdf

State Record Office (2012d), South Australian Government, RK051 - ARM - Disposal V1 http://government.archives.sa.gov.au/sites/default/files/20131106%20Adequate%20Records%20Management%20-%20Disposal%20Final%20V1_Copy.pdf

State Record Office (2014), Government of South Australia, General Disposal authority 34, viewed at http://government.archives.sa.gov.au/sites/default/files/20150302%20General%20Disposal%20Schedule%20No.%2034%20Final%20V2_Copy.pdf

VAILE, D. (2014). The Cloud and data sovereignty after Snowden. Australian Journal of Telecommunications and the Digital Economy, vol.2, no.31.1, pp.31.58. University of South Australia

Yusof, ZM & Chell, RW (1998), 'The eluding definitions of records and records management: is a universally acceptable definition possible? Part 2: Defining records management', Records ViManagement Journal, vol. 9, no. 1, pp. 9-20.

## Copyright